\documentclass[12pt]{article}
\usepackage{lineno}
\usepackage{graphicx}
\usepackage{amsmath}
\usepackage{hhline}
\usepackage{amssymb}
\usepackage{times}
\usepackage{cite}
\usepackage{array}
\usepackage{multirow}
\usepackage{changebar}
\usepackage{pifont}

\usepackage{color}
\definecolor{rltred}{rgb}{0.75,0,0}
\definecolor{rltgreen}{rgb}{0,0.5,0}
\definecolor{rltblue}{rgb}{0,0,0.75}

\newif\ifpdf
\ifx\pdfoutput\undefined
    \pdffalse          
\else
    \pdfoutput=1       
    \pdftrue
\fi

\ifpdf

\usepackage{thumbpdf}
\usepackage[pdftex,
        colorlinks=true,
        urlcolor=rltblue,       
        filecolor=rltgreen,     
        linkcolor=rltred,       
        pdftitle={Example File for pdflatex},
        pdfauthor={H1},
        pdfsubject={Template file},
        pdfkeywords={High-Energy Physics, Particle Physics},
        pdfpagemode=None,
        bookmarksopen=true]{hyperref}
 
\fi

\newlength{\dinwidth}
\newlength{\dinmargin}
\newcommand{\GeV}{{\rm GeV}}

\newcommand{\pb}{\rm pb}
\newcommand{\cm}{\rm cm}

\newcommand{\mean}[1]{\langle#1\rangle}
\newcommand{\degr}{^\circ}

\begin{document}

\newcommand{\pom}{{I\!\!P}}
\newcommand{\reg}{{I\!\!R}}
\newcommand{\slowpi}{\pi_{\mathit{slow}}}
\newcommand{\fiidiii}{F_2^{D(3)}}
\newcommand{\fiidiiiarg}{\fiidiii\,(\beta,\,Q^2,\,x)}
\newcommand{\n}{1.19\pm 0.06 (stat.) \pm0.07 (syst.)}
\newcommand{\nz}{1.30\pm 0.08 (stat.)^{+0.08}_{-0.14} (syst.)}
\newcommand{\fiidiiiful}{F_2^{D(4)}\,(\beta,\,Q^2,\,x,\,t)}
\newcommand{\fiipom}{\tilde F_2^D}
\newcommand{\ALPHA}{1.10\pm0.03 (stat.) \pm0.04 (syst.)}
\newcommand{\ALPHAZ}{1.15\pm0.04 (stat.)^{+0.04}_{-0.07} (syst.)}
\newcommand{\fiipomarg}{\fiipom\,(\beta,\,Q^2)}
\newcommand{\pomflux}{f_{\pom / p}}
\newcommand{\nxpom}{1.19\pm 0.06 (stat.) \pm0.07 (syst.)}
\newcommand {\gapprox}
   {\raisebox{-0.7ex}{$\stackrel {\textstyle>}{\sim}$}}
\newcommand {\lapprox}
   {\raisebox{-0.7ex}{$\stackrel {\textstyle<}{\sim}$}}
\def\gsim{\,\lower.25ex\hbox{$\scriptstyle\sim$}\kern-1.30ex%
\raise 0.55ex\hbox{$\scriptstyle >$}\,}
\def\lsim{\,\lower.25ex\hbox{$\scriptstyle\sim$}\kern-1.30ex%
\raise 0.55ex\hbox{$\scriptstyle <$}\,}
\newcommand{\pomfluxarg}{f_{\pom / p}\,(x_\pom)}
\newcommand{\dsf}{\mbox{$F_2^{D(3)}$}}
\newcommand{\dsfva}{\mbox{$F_2^{D(3)}(\beta,Q^2,x_{I\!\!P})$}}
\newcommand{\dsfvb}{\mbox{$F_2^{D(3)}(\beta,Q^2,x)$}}
\newcommand{\dsfpom}{$F_2^{I\!\!P}$}
\newcommand{\gap}{\stackrel{>}{\sim}}
\newcommand{\lap}{\stackrel{<}{\sim}}
\newcommand{\fem}{$F_2^{em}$}
\newcommand{\tsnmp}{$\tilde{\sigma}_{NC}(e^{\mp})$}
\newcommand{\tsnm}{$\tilde{\sigma}_{NC}(e^-)$}
\newcommand{\tsnp}{$\tilde{\sigma}_{NC}(e^+)$}
\newcommand{\sst}{$\star \star$}
\newcommand{\ssst}{$\star \star \star$}
\newcommand{\sssst}{$\star \star \star \star$}
\newcommand{\tw}{\theta_W}
\newcommand{\sw}{\sin{\theta_W}}
\newcommand{\cw}{\cos{\theta_W}}
\newcommand{\sww}{\sin^2{\theta_W}}
\newcommand{\cww}{\cos^2{\theta_W}}
\newcommand{\trm}{m_{\perp}}
\newcommand{\trp}{p_{\perp}}
\newcommand{\trmm}{m_{\perp}^2}
\newcommand{\trpp}{p_{\perp}^2}
\newcommand{\alp}{\alpha_s}

\newcommand{\alps}{\alpha_s}
\newcommand{\alpsmz}{\alpha_s(m_Z)}
\newcommand{\alpomi}{\alpha_0(\mu_I=2\GeV)}
\newcommand{\lqcd}{\Lambda_\mathrm{QCD}}
\newcommand{\diff}{\mathrm{d}}
\def\MSbar{$\overline{\mbox{\rm MS}}$}
\newcommand{\sqrts}{$\sqrt{s}$}
\newcommand{\LO}{$O(\alpha_s^0)$}
\newcommand{\Oa}{$O(\alpha_s)$}
\newcommand{\Oaa}{$O(\alpha_s^2)$}
\newcommand{\PT}{p_{\perp}}
\newcommand{\JPSI}{J/\psi}
\newcommand{\sh}{\hat{s}}
\newcommand{\uh}{\hat{u}}
\newcommand{\MP}{m_{J/\psi}}
\newcommand{\PO}{I\!\!P}
\newcommand{\xbj}{x}
\newcommand{\xpom}{x_{\PO}}
\newcommand{\ttbs}{\char'134}
\newcommand{\xpomlo}{3\times10^{-4}}  
\newcommand{\xpomup}{0.05}  
\newcommand{\dgr}{^\circ}
\newcommand{\pbarnt}{\,\mbox{{\rm pb$^{-1}$}}}
\newcommand{\gev}{\,\mbox{GeV}}
\newcommand{\WBoson}{\mbox{$W$}}
\newcommand{\fbarn}{\,\mbox{{\rm fb}}}
\newcommand{\fbarnt}{\,\mbox{{\rm fb$^{-1}$}}}

\newcommand{\hch}{{\scriptscriptstyle h \in {\rm CH}}}
\newcommand{\ich}{{\scriptscriptstyle i \in {\rm CH}}}
\newcommand{\ee}{\mbox{$e^+e^-$~}}
\newcommand{\mf}{\mu_{\scriptscriptstyle F}}
\newcommand{\mi}{\mu_{\scriptscriptstyle I}}
\newcommand{\mr}{\mu_{\scriptscriptstyle R}}
\newcommand{\mrad}{\,\textrm{mrad}}

%
%
\newcommand{\qsq}{\ensuremath{Q^2} }
\newcommand{\gevsq}{\ensuremath{\mathrm{GeV}^2} }
\newcommand{\et}{\ensuremath{E_t^*} }
\newcommand{\rap}{\ensuremath{\eta^*} }
\newcommand{\gp}{\ensuremath{\gamma^*}p }
\newcommand{\dsiget}{\ensuremath{{\rm d}\sigma_{ep}/{\rm d}E_t^*} }
\newcommand{\dsigrap}{\ensuremath{{\rm d}\sigma_{ep}/{\rm d}\eta^*} }
\def\Journal#1#2#3#4{{#1} {\bf #2} (#3) #4}
\def\NCA{\em Nuovo Cimento}
\def\NIM{\em Nucl. Instrum. Methods}
\def\NIMA{{\em Nucl. Instrum. Methods} {\bf A}}
\def\NPB{{\em Nucl. Phys.}   {\bf B}}
\def\PLB{{\em Phys. Lett.}   {\bf B}}
\def\PRL{\em Phys. Rev. Lett.}
\def\PRD{{\em Phys. Rev.}    {\bf D}}
\def\ZPC{{\em Z. Phys.}      {\bf C}}
\def\EJC{{\em Eur. Phys. J.} {\bf C}}
\def\CPC{\em Comp. Phys. Commun.}

\begin{titlepage}

\noindent
\begin{flushleft}
DESY 05-225 \hfill ISSN 0418-9833 \\
December 2005
\end{flushleft}


\vspace{2cm}

\begin{center}
\begin{Large}

{\bf Measurement of Event Shape Variables in Deep-Inelastic Scattering at HERA}

\vspace{2cm}

H1 Collaboration

\end{Large}
\end{center}

\vspace{2cm}

\begin{abstract} \noindent
  Deep-inelastic $ep$ scattering data taken with the H1 detector at HERA and
  corresponding to an integrated luminosity of $106~\pb^{-1}$ are used to study
  the differential distributions of event shape variables. These include
  thrust, jet broadening, jet mass and the $C$-parameter. The four-momentum
  transfer $Q$ is taken to be the relevant energy scale and ranges between 
  $14~\GeV$ and $200~\GeV$. The event shape distributions are compared with
  perturbative QCD predictions, which include resummed contributions
  and analytical power law corrections, the latter accounting for non-perturbative
  hadronisation effects. The data clearly exhibit the running of the 
  strong coupling $\alp(Q)$ and are consistent with a universal  power correction
  parameter $\alpha_0$ for all event shape variables. A combined QCD fit using all
  event shape variables yields 
  $\alps(m_Z) = 0.1198 \pm 0.0013\ ^{+0.0056} _{-0.0043}$ and 
  $\alpha_0 = 0.476 \pm  0.008 \ ^{+0.018} _{-0.059}$.
\end{abstract}

\vspace{1.5cm}

\begin{center}
  Submitted to Eur. Phys. J. C
\end{center}

\end{titlepage}

\begin{flushleft}

A.~Aktas$^{9}$,                
V.~Andreev$^{25}$,             
T.~Anthonis$^{3}$,             
B.~Antunovic$^{26}$,           
S.~Aplin$^{9}$,                
A.~Asmone$^{33}$,              
A.~Astvatsatourov$^{3}$,       
A.~Babaev$^{24}$,              
S.~Backovic$^{30}$,            
J.~B\"ahr$^{38}$
% ...

\bigskip{\it
% ...
$ ^{3}$ Inter-University Institute for High Energies ULB-VUB, Brussels;
          Universiteit Antwerpen, Antwerpen; Belgium$^{ c}$ \\
% ...
$ ^{9}$ DESY, Hamburg, Germany \\
% ...
$ ^{24}$ Institute for Theoretical and Experimental Physics,
          Moscow, Russia$^{ k}$ \\
 $ ^{25}$ Lebedev Physical Institute, Moscow, Russia$^{ e}$ \\
 $ ^{26}$ Max-Planck-Institut f\"ur Physik, M\"unchen, Germany \\
% ...
$ ^{30}$ Faculty of Science, University of Montenegro,
          Podgorica, Serbia and Montenegro$^{ e}$ \\
% ...
$ ^{33}$ Dipartimento di Fisica Universit\`a di Roma Tre
          and INFN Roma~3, Roma, Italy \\
% ...
$ ^{38}$ DESY, Zeuthen, Germany \\
% ...
$ ^c$ Supported by FNRS-FWO-Vlaanderen, IISN-IIKW and IWT
      and  by Interuniversity
Attraction Poles Programme,
      Belgian Science Policy \\
% ...
$ ^e$ Supported by the Deutsche Forschungsgemeinschaft \\
% ...
$ ^k$ Partially Supported by Russian Foundation
      for Basic Research,  grants  03-02-17291
      and  04-02-16445 \\
}
\end{flushleft}

\newpage

\section{Introduction}
\label{introduction}
This paper presents a study of the hadronic final state in deep-inelastic scattering (DIS),
 $ep\to eX$, using event shape variables.
Event shapes probe the energy-momentum flow and  are sensitive to perturbative QCD (the hard scattering) 
and non-perturbative QCD (hadronisation).
Compared to some other hadronic final state observables,
such as jet cross sections, which employ a limited part of the phase space only,
the event shape variables include the full statistics and,
by definition, are rather insensitive 
to hadronic energy scale uncertainties.
With DIS at HERA a wide range of the scale $Q$ is available for analysis in a single
experiment, where $Q$ is the four-momentum transferred by the exchanged boson from the
electron to the proton.

Studies of event shapes require an appropriate treatment of
hadronisation effects, which often are corrected for using phenomenological
models built into event generators.
Alternatively, these
effects have been treated by power corrections  $\mathcal{O}(1/Q)$
as calculated analytically from first principles by
extending perturbative methods into the non-perturbative
regime~\cite{Dokshitzer:1995zt,Dokshitzer:1996qm,Dokshitzer:1997ew}.
Within this framework, the event shapes are described
by the strong coupling constant $\alpha_s$ and an effective
coupling parameter $\alpha_0$ for the hadronisation corrections.
Measurements of event shapes thus represent a sensitive test
of the power correction approach, and they allow $\alpha_s$
to be determined.

Previous analyses of mean values of event shape variables
in DIS, as performed by the H1~\cite{Adloff:1999gn} and ZEUS~\cite{Chekanov:2002xk} collaborations at HERA,
support the power correction approach. An observed large spread of
values obtained for $\alpha_s$, however, indicates that
higher-order QCD corrections cannot be neglected. Support for
power corrections as an appropriate description of the
hadronisation is also provided  by analyses of mean
values and distributions of event shape variables
measured in \ee experiments~\cite{MovillaFernandez:2001ed,Abdallah:2002xz,Heister:2003aj} . 

Early studies of event shape distributions in deep inelastic
scattering revealed that the fixed-order QCD calculations
available at the time
were insufficient for describing the data~\cite{unknown:1998kr}. 
Higher-order corrections have since become available in the form
of soft gluon resummed calculations matched to NLO matrix
elements~\cite{Dasgupta:2002dc}. This puts the study of event shape distributions
and of the interplay between perturbative and non-perturbative
QCD in the description of the hadronic final state on a new
quantitative level.

In this paper the distributions of five event shape variables are studied.
A QCD analysis is performed based on the resummed and matched calculations supplemented by power corrections.
This leads to determinations of the power correction
coefficient $\alpha_0$ and of $\alpha_s$ and its dependence
on the scale $Q$.

\section{Event Shape Variables}
\label{eventshapes}

The aim of this analysis is to study event shape variables within QCD as a
function of the relevant hard scale, which in DIS is taken
to be the four-momentum $Q$ of the exchanged boson.
In addition to the hadronic flow from the hard scattering there are also
hadrons in the final state stemming from 
the proton dissociation, which occurs at much lower scales of about the proton mass.
Therefore it is necessary to separate this proton remnant from the hard scattering part of the event.

\subsection{The Breit Frame}
\label{breitframe}

In the quark parton model the separation of the proton remnant from the hard scattering
is clearest in the Breit frame of 
reference, defined by $2x\vec p + \vec q=0$, where $x$ is the Bjorken scaling 
variable, $\vec p$ the momentum of the proton and $\vec q$ the momentum of the 
exchanged boson.
The $z$~axis in the Breit frame is defined to coincide with the proton-boson 
axis, the proton moving in the $+z$~direction.
Particles from the remnant are almost collinear to the proton direction, hence 
the hemisphere defined by pseudorapidity\footnote{The pseudorapidity 
  is defined as $\eta=-\ln \tan (\theta/2)$ with $\theta$ 
  the polar angle with respect to the $z$ axis.} 
$\eta>0$ is labelled the remnant hemisphere.
In contrast, in the quark parton model the struck quark populates only the current hemisphere ($\eta<0$).

Higher order processes generate transverse momenta in the final state
and may even lead to particles from the hard subprocess leaking into
the remnant hemisphere.
Still, without know\-ledge of the detailed structure of the hadronic final state, 
the Breit frame allows for optimal separation of the current region from 
the proton remnant.
All event shapes referred to in this paper are defined using particles in the 
current hemisphere only.

The kinematic quantities needed to perform the Breit frame transformation are
calculated using the electron-sigma method~\cite{Bassler:1994uq,Bassler:1997tv}.
The virtuality $Q^2$ of the exchanged boson is reconstructed using the energy and polar 
angle of the scattered electron, and
the inelasticity $y$ is determined employing in addition the energy and longitudinal momentum of all
hadronic objects measured in the laboratory frame.
This method results in good experimental resolution and is relatively insensitive to initial state QED bremsstrahlung.

\subsection{Definition of Event Shape Variables}
\label{definitions}

The event shape variables studied in this paper are those for which the calculations
of power corrections are available.
They are defined as follows.

\par\noindent
The {\em Thrust} variable $\tau$\/ measures the longitudinal momentum
components projected onto the boson axis. It is defined as 
\begin{eqnarray}
\tau=1-T\quad\mathrm{with}\quad  T&=&\frac{\sum_h|\vec p_{z,h}|}{\sum_h|\vec p_h|} \ .
  \label{eqn:thrust} 
\end{eqnarray}
The variable $\tau_C$\/ calculates the {\em Thrust} with respect to the direction $\vec n_T$
which maximises the sum of the longitudinal momenta of all particles
in the current hemisphere along this axis. It is defined as
\begin{eqnarray}
\tau_C=1-T_C\quad\mathrm{with}\quad T_C&=&\max_{\vec n_T} \frac{\sum_h|\vec p_h\cdot \vec n_T|}{\sum_h|\vec p_h|} \ .
  \label{eqn:thrustc}
\end{eqnarray}
The {\em Jet Broadening $B$}\/ measures the scalar sum of transverse
momenta with respect to the boson axis
\begin{eqnarray}
  B&=&\frac{\sum_h|\vec p_{t,h}|}{2\sum_h|\vec p_h|} \ .
  \label{eqn:bparameter}
\end{eqnarray}
The squared {\em Jet Mass $\rho$}\/ is normalised to four times the
squared scalar momentum sum in the current hemisphere
\begin{eqnarray}
  \rho&=&\frac{(\sum_h E_h)^2 - (\sum_h\vec p_h)^2}{(2\sum_h|\vec p_h|)^2}.
  \label{eqn:jetmass} 
\end{eqnarray}
In the following the symbol $\rho_0$ is used, which indicates that
in the above definition the hadrons are treated as massless, replacing
the energy $E_h$ by the modulus of the 3-momentum $|\vec p_h|$.

The {\em $C$-Parameter}\/ is defined as
\begin{eqnarray}
  C&=&\frac{3}{2}\frac{\sum_{h,h'}|\vec p_h||\vec p_{h'}|\cos^2\theta_{hh'}}
                      {(\sum_h|\vec p_h|)^2} \ ,
  \label{eqn:cparameter}
\end{eqnarray}
where $\theta_{hh'}$ is the angle between particles $h$ and $h'$.

In Eq.~\ref{eqn:thrust}-\ref{eqn:cparameter} the momenta are
 defined in the Breit frame and the sums extend
over all particles in the current hemisphere.

An event is only accepted if the energy in the current hemisphere exceeds some 
value $\epsilon_{\mathrm{lim}}$.
This is necessary to ensure the infrared and collinear safety of the
observables, because higher order processes may lead to event configurations
in which the partons are scattered into the remnant hemisphere and the current
hemisphere is completely empty, except for arbitrarily soft emissions. 
In the analysis the events are required to fulfill the condition
\begin{eqnarray}
  \sum\limits_h E_h & > & \epsilon_{\mathrm{lim}}=Q/10\,,
  \label{eqn:echcut} 
\end{eqnarray}
as part of the event shape definitions.
The precise value of this cut-off turns out not to be crucial~\cite{Adloff:1999gn}. 

The event shape variables may be distinguished according to the event axis used.
The definitions of $\tau$ and $B$ employ momentum vectors projected onto the
boson direction, while the others do not, like their counterparts in $e^+e^-$ 
reactions.
Explicit use of the boson direction implies sensitivity to radiation into the 
remnant hemisphere through recoil effects on the current 
quark~\cite{Dasgupta:2002dc}.

Throughout the paper the symbol $F$ will be used as a generic name for any
event shape variable.  
Note that for all variables $F$ tends to zero in the case of quark parton model 
reactions neglecting hadronisation effects (small values of $F$ correspond
to pencil like configurations of the hadronic final state).
Theoretical calculations of event shape distributions and means are 
discussed in section~\ref{qcdanalysis}.

\section{Experimental Technique}

The data were collected with the H1 detector at HERA during the years 1995 -- 2000 and 
correspond to an integrated luminosity of $\mathcal{L}_{\textrm{int}}=106~\pb^{-1}$.
The collider was operated with electrons or positrons of 
$E_e=27.6\,\GeV$ and protons of $E_p=820\,\GeV$ or $E_p=920\,\GeV$,
yielding centre-of-mass energies $\sqrt{s}$ of $301~\GeV$ and
$319~\GeV$, respectively. 
For the present study, 
three data samples are used:
\begin{itemize}
\item \ \ $e^+p$, \ $\sqrt{s}\simeq 301~\GeV$,
  $\mathcal{L}_{\textrm{int}}=30~\pb^{-1}$ (1995-1997) ;
\item \ \ $e^-p$, \ $\sqrt{s}\simeq 319~\GeV$,
  $\mathcal{L}_{\textrm{int}}=14~\pb^{-1}$ (1998-1999) ;
\item \ \ $e^+p$, \ $\sqrt{s}\simeq 319~\GeV$,
  $\mathcal{L}_{\textrm{int}}=62~\pb^{-1}$ (1999-2000) .
\end{itemize}

The identification of neutral current DIS events is based on the reconstruction
of an event vertex and of the scattered electron\footnote{
Here and in the following ``electron'' is used to refer to both electron and positron unless explicitly  stated otherwise.
} in the central tracker and the
calorimeter.

\subsection{H1 Detector}

A detailed description of the H1 detector can be found 
in~\cite{Abt:1997hi,Abt:1997xv}.
The most important detector components for the present analysis are the 
liquid argon (LAr) calorimeter and the central tracking system.
H1 uses a right-handed coordinate system with the $z$~axis along
the beam direction, the $+z$ or ``forward" direction being that of the outgoing proton beam.
The polar angle $\theta$ is defined with respect to the $z$ axis.

The LAr sampling calorimeter ($4\degr \le \theta \le 154\degr$) consists of 
lead/liquid argon electromagnetic sections and stainless steel/liquid argon sections for the 
measurement of hadronic energy.
An {\em in situ}\/ calibration provides energy scales.
The electron energy scale uncertainty in the LAr calorimeter varies between $0.7\%$ 
 and $3\%$~\cite{Adloff:2003uh}.
The hadronic energy measurement is performed by applying a weighting technique
to the electromagnetic and the hadronic components of the energy deposition in
order to account for the unequal response to electrons and hadrons.  
The systematic uncertainty on the hadronic energy scale amounts to $2\%$.
In the backward region ($153\degr \le \theta \le 177\degr$) energy is detected
by a lead/scintillating fibre Spaghetti-type calorimeter~\cite{Appuhn:1997na}.

The central tracking system ($25\degr \le \theta \le 155\degr$) is located inside
 the LAr calorimeter  and consists of drift and proportional chambers. 
The chambers and calorimeters are surrounded by a superconducting
solenoid providing a uniform field of $1.15\,\mathrm{T}$ inside the tracking
volume. 
The scattered electron is identified  by associating tracking
information with the corresponding electromagnetic cluster in the 
LAr calorimeter. 
The electron scattering angle is known within $3\,\mathrm{mrad}$.

For the present analysis the hadronic final state is reconstructed from combined
objects, built from calorimeter clusters and tracks, 
using an energy flow algorithm which
ensures that no double counting of energy occurs.
Compared to clusters alone the combined objects improve the reconstruction of
low momentum particles.

\subsection{Event Selection}

Several quality cuts are applied to the data.
A scattered electron
has to be found in the LAr calorimeter with a reconstructed energy $E'_{e}$ exceeding  
$11~\GeV$, which ensures a trigger efficiency above 98\%.
The $z$~position of the event vertex has to be reconstructed within $\pm 35\,\cm$ 
of the nominal position of the interaction point, which reduces contributions 
from beam induced background.
Non-$ep$ background is further reduced by requiring an event timing which 
matches the HERA bunch crossing.
To suppress badly measured events the missing transverse momentum has to be 
below $15~\GeV$.
The total longitudinal energy balance must satisfy $40~\GeV < \sum_i
E_i(1-\cos \theta_i)  < 70~\GeV$, where the sum runs over all detected particles.
This reduces photoproduction background and
initial state photon radiation. 
Such QED radiative effects, backgrounds and poorly reconstructed events are
 further suppressed by demanding that
the calorimetric energy measurement of the lepton be consistent within 
$10\% $ with that derived from the double angle method~\cite{Hoeger:1991wj,Bentvelsen:1992fu}.

The kinematic region covered by the analysis is defined by ranges of boson
virtuality $Q^2$ and inelasticity $y$:
$$ 196<Q^2<40,000~\GeV^2 \ ,$$
$$0.1<y<0.7 \ ,$$
which are reconstructed using the electron-sigma method~\cite{Bassler:1994uq,Bassler:1997tv}.

In total about $108,000$ events satisfy the selection criteria.
The PYTHIA~\cite{Sjostrand:2000wi} program is used to estimate the background from photoproduction events,
in which a hadron in the LAr calorimeter is misidentified as an electron.
This background is found to be negligible in all bins of the various event shape distributions.

\subsection{Correction to the Hadron Level}
The correction of the data for limited detector acceptance uses the simulation program
 RAPGAP~2.8 \cite{Jung:1995gf} with parton showers and string fragmentation as implemented in 
JETSET~\cite{Sjostrand:1994yb}. 
The parton density functions (pdfs) of the proton are taken from the 
CTEQ5L~\cite{Lai:1999wy} set.
The data are corrected for QED radiation effects using the HERACLES~\cite{Kwiatkowski:1992es} program.
Bin-to-bin correction factors are determined from the Monte Carlo event samples
 passed through a detailed simulation of the H1
detector and subjected to the same reconstruction and analysis chain
as the data.

The effects of limited detector resolution are corrected for in a separate step.
In the bins used for the QCD analysis presented in section~\ref{qcdanalysis}
 the purities are typically $30-50 \%$.
To correct
for the corresponding bin correlations the covariance matrix is
determined with an iterative Bayesian unfolding method
\cite{D'Agostini:1995zf}. 
The data are unfolded to the level of hadrons in order to compare
with resummed calculations supplemented by power corrections for hadronisation.
The correction procedure is performed separately for the three data 
samples.

\subsection{Experimental Uncertainties}
\label{detcorrectionsection}

Several studies are carried out to estimate the experimental systematic error by
using alternative settings or assumptions in the Monte Carlo programs.
The correlations of these changes between bins 
of an event shape variable are taken into account unless stated otherwise. 

The event kinematics and therefore the boost to the Breit frame of reference
depends strongly on the momentum of the reconstructed electron.
Thus the electromagnetic energy scale is varied by its uncertainty of
$\pm(0.7\%-3\%)$, depending on the 
$z$ position of the electron cluster within the LAr calorimeter.
In addition, the polar and azimuthal angles of the electron are changed by $
\pm 3\mrad$ each. 
  
The event shape variables are by definition insensitive to
variations of the overall hadronic energy scale. 
The effects of the uncertainty on the hadronic 
intercalibration are
investigated by shifting the calibration constants of
neighbouring calorimeter regions with respect to each other~\cite{Kluge:2004ef}.
Since the final event shape distributions average over all calorimeter 
regions, the resulting uncertainty on the event shape variables is small. 

The model dependence of the correction procedure is estimated by replacing RAPGAP with the
DJANGOH~1.2~\cite{Charchula:1994kf} event generator, which employs the colour dipole model of
ARIADNE~\cite{Lonnblad:1992tz} to simulate higher order QCD radiation.
The model uncertainty is estimated as the difference between the results 
obtained with the two Monte Carlo samples.
One half of this is treated as uncorrelated between the bins.
The other half is assumed to be fully correlated between bins.
 
An estimate of the possible intrinsic bias from the unfolding procedure is obtained
by unfolding Monte Carlo event samples with themselves. 
The residuum between the correct hadron level and the unfolded result is taken
as the unfolding error, which is assumed not to be correlated between
the bins of the distributions.

All systematic errors from different sources are added in
quadrature, the model uncertainty and the
electromagnetic energy scale uncertainty being the largest individual
contributions.

\subsection{Combination of Data Sets}
 
The combination of separately unfolded data sets proceeds in two steps, first combining the
positron data from the two different centre-of-mass energies  and second combining the
positron and electron data.
The distributions of the two positron data sets are compatible with each other
within errors. 
They are combined by calculating 
the luminosity weighted averages for all bins of the distributions.

In general, the event shape distributions of the $e^+$ and $e^-$ data are in very
good agreement with each other, though some discrepancies are observed at the
highest $Q$ scales.
Differences in the event shape distributions between the $e^+$ and $e^-$  data sets are
expected because of $Z$ exchange contributions to the cross section~\cite{privsalam}. 
Unfortunately, only the $\gamma$ exchange component is accounted for in the
calculations of event shape distributions which are fitted to the data.
The  $e^+$ and $e^-$ event shape distributions are thus averaged, weighted with the
corresponding cross section such that most of the $Z$ contribution cancels. 
Due to the smaller integrated luminosity, the $e^-$ data increase the
statistical error of the final spectra. 
However, this effect is partly compensated by the larger cross section for $e^-p$
scattering at high $Q$.

The resulting  mean values of $Q$ and $x$ are slightly modified by the
combination procedure, giving the values listed in Table~\ref{xQtable}.
The average centre-of-mass energy of the combined set is $\sqrt{s}=316~\GeV$.

\begin{table}[htb]
  \begin{center}
    \begin{tabular}{|c||l|l|l|l|l|l|l|}
      \hline
      \# of $Q$ bin &1&2&3&4&5&6&7\\
      \hline
      $Q$ Interval$/\GeV$&[14,16]&[16,20]&[20,30]&[30,50]&[50,70]&[70,100]&[100,200]\\
      \hline
      $\mean{Q}/\GeV$&\multicolumn{1}{c|}{14.9}& \multicolumn{1}{c|}{17.7}&\multicolumn{1}{c|}{ 23.8}&\multicolumn{1}{c|}{ 36.9}&\multicolumn{1}{c|}{ 57.6}&\multicolumn{1}{c|}{ 80.6}&\multicolumn{1}{c|}{115.6}\\
      \hline
      $\mean{x}$&\multicolumn{1}{c|}{0.00841}&\multicolumn{1}{c|}{0.0118}&\multicolumn{1}{c|}{0.0209}&\multicolumn{1}{c|}{0.0491}&\multicolumn{1}{c|}{0.116}&\multicolumn{1}{c|}{0.199}&\multicolumn{1}{c|}{0.323}\\
      \hline
    \end{tabular}
  \end{center} 
  \caption{$Q$ intervals, mean $Q$ and mean Bjorken $x$ for the seven 
    $Q$ bins of the analysis.}
  \label{xQtable}  
\end{table}

\section{Event Shape Measurements}
\label{resultssec}

\begin{figure} 
  \begin{center}
    \vspace{-1cm}
    \includegraphics[width=0.98\textwidth]{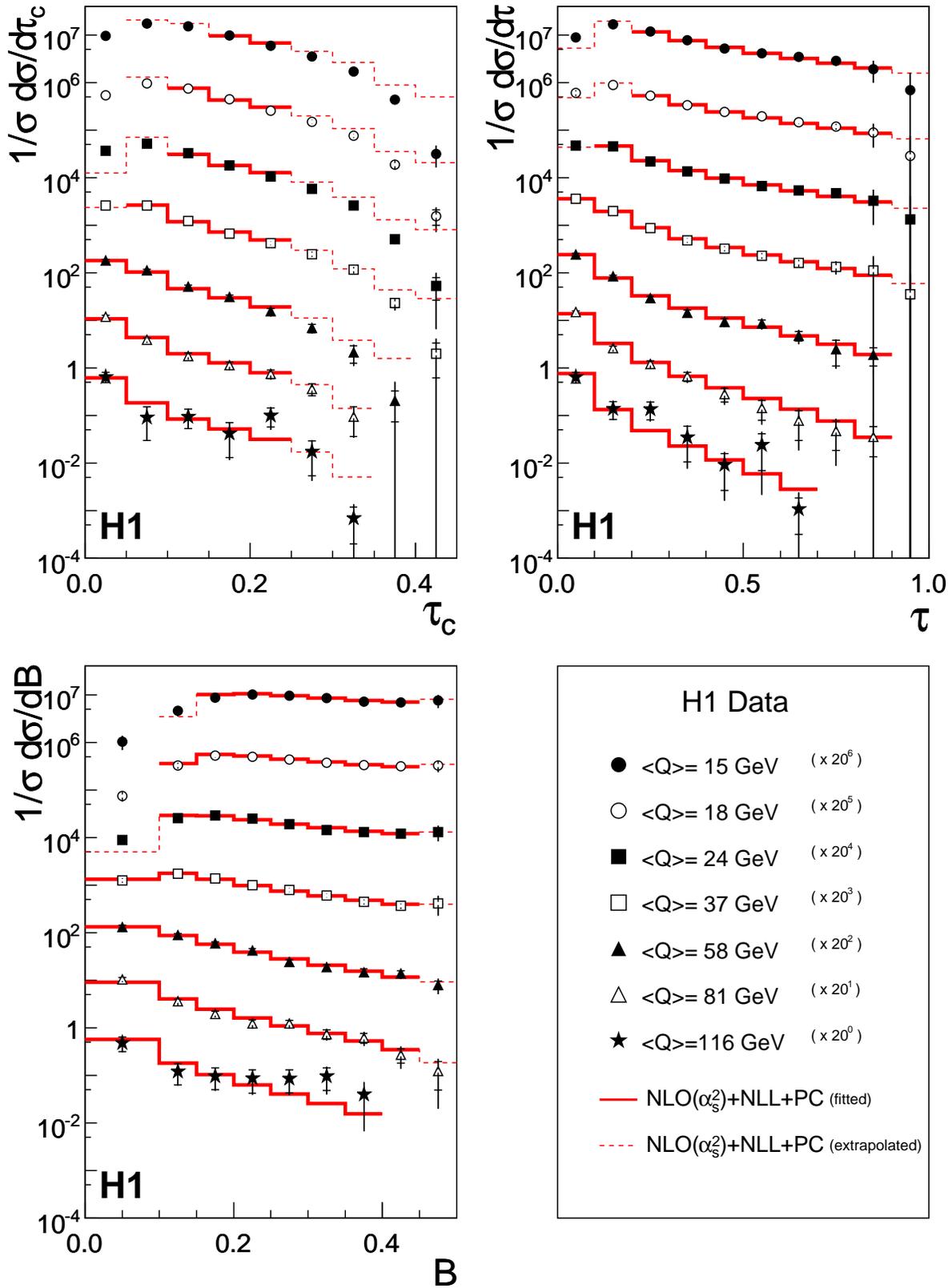}
    \vspace{-1cm}
  \end{center}
  \caption{ Normalised event shape distributions corrected to the hadron level
    for $\tau_C$, $\tau$ and $B$.
    The data are presented with statistical errors (inner bars)
    and total errors (outer bars).
    The measurements are compared with fits based on a NLO QCD calculation including 
    resummation (NLL) and supplemented by power corrections (PC).
    The fit results are shown as solid lines and are extended as
    dashed lines to those data points which are not included in the QCD fit 
    (see Section~\ref{fitstospectra}).}
  \label{figureshapesdistributions1}
\end{figure}

\begin{figure} [htb]
  \begin{center}
    \vspace{-1cm}
    \includegraphics[width=\textwidth]{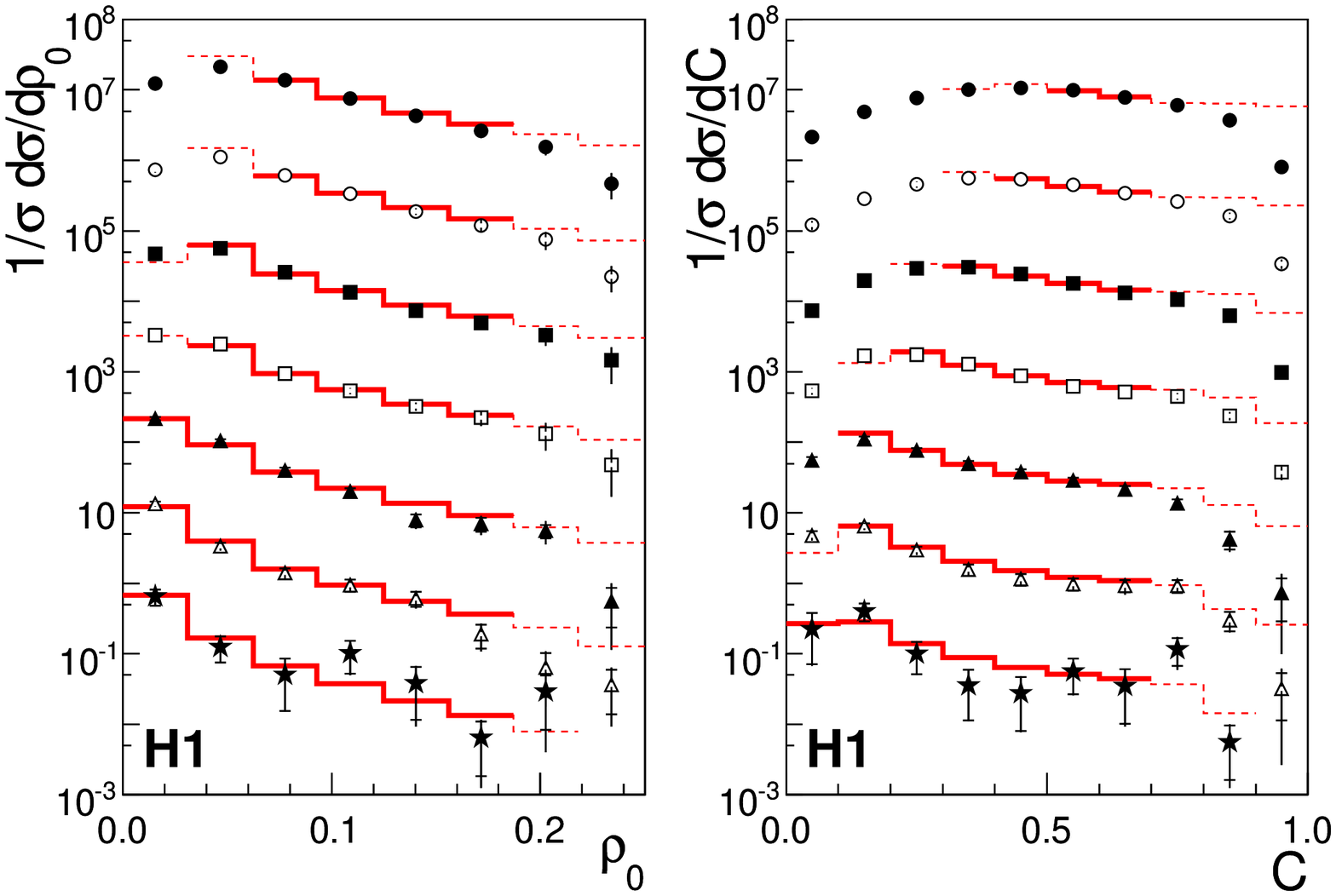}
    \vspace{-1cm}
  \end{center}
  \caption{ Normalised event shape distributions corrected to the hadron level
    for $\rho_0$ and the $C$-parameter. 
    The data are presented with statistical errors (inner bars)
    and total errors (outer bars).
    The measurements are compared with fits based on a NLO QCD calculation including 
    resummation (NLL) and supplemented by power corrections (PC).
    The fit results are shown as solid lines and are extended as
    dashed lines to those data points which are not included in the QCD fit (see Section~\ref{fitstospectra}).
   The symbols and scale factors are defined in Fig.~\ref{figureshapesdistributions1}.}
\label{figureshapesdistributions2}
\end{figure}

The normalised event shape distributions at the hadron level for thrust, jet broadening, jet mass
and the $C$-parameter are shown in Figs.~\ref{figureshapesdistributions1} and
\ref{figureshapesdistributions2} over a wide range of 
$\langle Q \rangle = 15 - 116~\GeV$.
The data points represent integrals over the
bins not applying bin centre corrections.
Except for the highest $Q$ bins, 
the precision of the measurements is not statistically limited.

For each variable the shape of the spectra changes considerably with increasing
$Q$, becoming narrower and evolving towards low values.
The strong $Q$ dependence of these spectra is characteristic of QCD.  
The results of the fits described in detail below are shown for comparison.

\begin{figure} 
  \begin{center}
    \includegraphics[width=.9\textwidth]{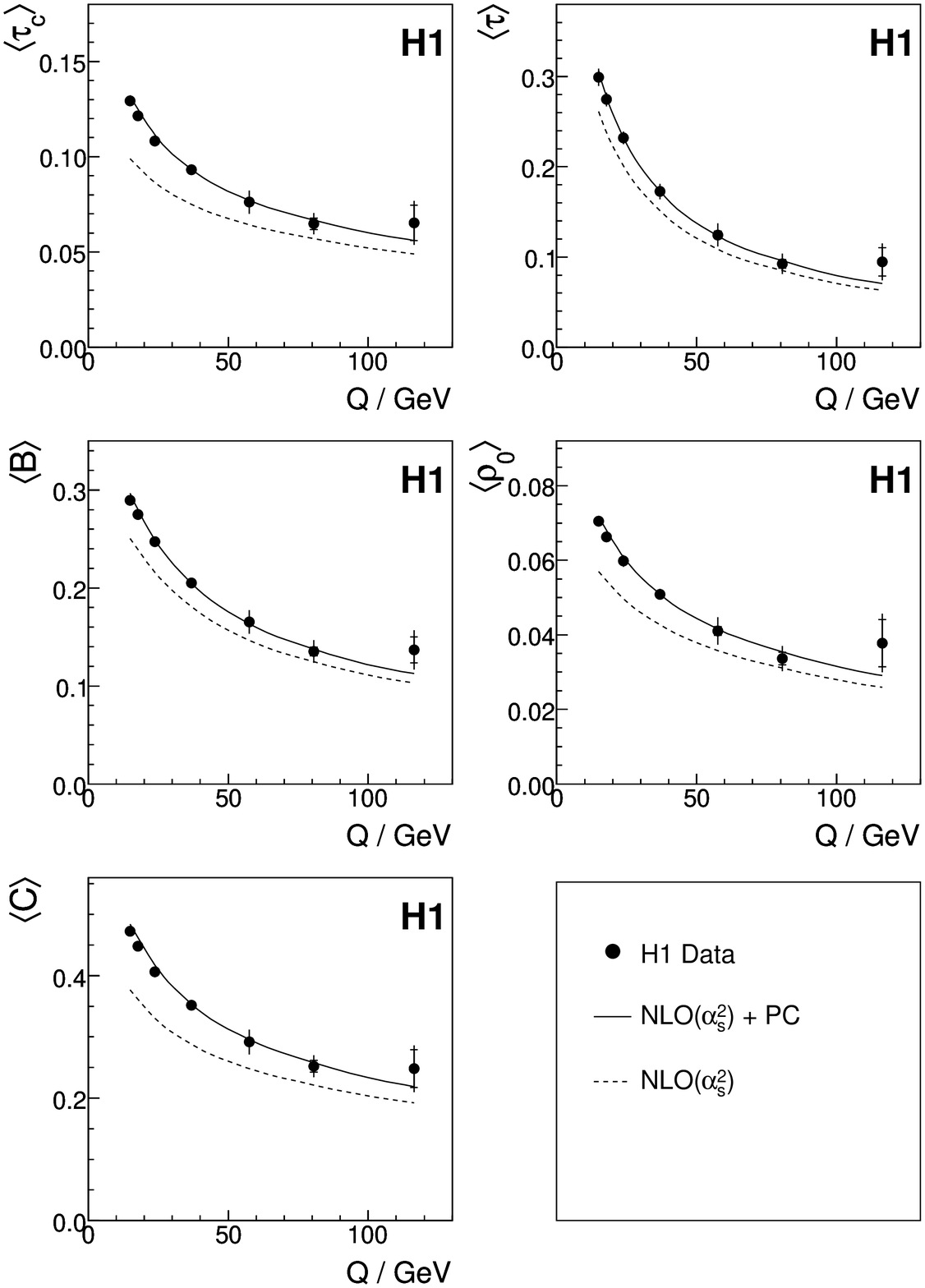}
  \end{center} 
  \caption{Mean values of event shape variables corrected to the hadron level as a function of the scale
    $Q$.
    The data, presented with statistical errors (inner bars)
    and total errors (outer bars),
    are compared with  the results of NLO QCD fits including power corrections (PC).
 The dashed curves show the NLO QCD contribution to the fits. }
\label{distrmeans}  
\end{figure}

The present paper focuses on differential distributions of the event shape variables.
However, in order to allow for a comparison to previous analyses 
 the mean values of these variables are also determined.
The results are shown in Fig.~\ref{distrmeans} as a function of $Q$. 
A steady decrease of the means with rising $Q$ is observed.
The results of the present analysis are in agreement with those  
previously obtained~\cite{Adloff:1999gn}.

\section{QCD Analysis}
\label{qcdanalysis}

\subsection{Phenomenology} \label{qcdpheno}

The QCD calculations used here contain a perturbative part (pQCD)
dealing with partons and in addition use power corrections (PC)
to describe the hadronisation.   
The perturbative part is made up of two contributions:
fixed order terms calculated to next-to-leading order (NLO) in the strong coupling constant and
resummed terms in the next-to-leading-logarithmic (NLL) approximation.

The NLO contribution consists of the first two terms of the perturbative
expansion in $\alp$ and, for any event shape variable $F$, has
the form 
\begin{equation}
  \frac{1}{\sigma}\frac{\diff\sigma^{\mathrm{NLO}}}{\diff F} = 
  c_1(F,Q)\alpha_S(\mu_r)+c_2(F,Q)\alpha_S^2(\mu_r),
\end{equation}
where $\mu_r$ is the renormalisation scale, chosen to be $Q$.
In order for the truncated series to be a good approximation to the exact
solution, $\alp(Q)$ needs to be small, i.e.\ $Q$ should be large. 

The coefficients $c_1$ and $c_2$ can be calculated from the matrix elements of
the hard scattering and the parton density functions of the proton. 
The fixed order coefficients  in the \MSbar\ scheme are
determined here using DISASTER++ \cite{Graudenz:1997gv} together with   
 DISPATCH~\cite{Dasgupta:2002dc}.
The parton density functions of the proton are taken from CTEQ5M1~\cite{Lai:1999wy}.

The region of $F\ll 1$ is dominated by events with soft and/or collinear parton emissions, leading
to large perturbative coefficients~\cite{ellisstirling}
\begin{equation}
  c_n(F) \underset{F \to 0}{\sim} \frac{\log^{2n-1}F}{F} \ . 
\end{equation}
In this region of low $F$ where the bulk of the data lies,
 the large logarithms need to be resummed to all orders in $\alp$.
For the event shape variables  considered the resummation has been performed by
Dasgupta and Salam~\cite{Dasgupta:2002dc,Dasgupta:2001eq} and is
available in the DISRESUM package.

To obtain a good description of the data over the full range of $F$, it is necessary to
add the fixed order and the resummed calculations, and to subtract any terms which
are counted twice, namely the $\mathcal{O}(\alp)$ and
$\mathcal{O}(\alp^2)$ terms of the resummed result. 
There are several valid matching schemes available.
Comparing the results from the different schemes leads to 
a residual ambiguity, which is considered in the uncertainty on the prediction. 
For the central values of the present fits the modified logR matching scheme is chosen, 
with the parameter $p_{\rm PT}$ set to two~\cite{Dasgupta:2002dc}.

All event shape variables are subject to non-per\-tur\-ba\-ti\-ve effects due to
hadronisation, for which calculations based on partons need to be corrected. 
The corrections can be determined by using fragmentation models as applied in
Monte Carlo programs. 
For many applications these models lead to a reasonable description of the data.
However, the interface between perturbative and non-per\-tur\-ba\-ti\-ve processes is
not well defined within these models, and there are phenomenological parameters to be tuned.

An alternative approach has been developed~\cite{Dokshitzer:1996qm} based on the observation
 that non-per\-tur\-ba\-ti\-ve corrections are in
general suppressed by powers of $(1/Q)$.
For the present event shape variables the leading corrections are proportional 
to $1/Q$ according to~\cite{Dokshitzer:1996qm}.
The power law behaviour of the corrections can be described by introducing an
 effective coupling $\alpha_\mathrm{eff}$, which is valid for low scales. 
In the perturbative region the effective coupling has to coincide with the
renormalised coupling $\alp(Q)$.
This is conventionally achieved by matching  $\alpha_\mathrm{eff}$ to $\alp$
at a scale $\mu_I = 2\,\GeV$.  
This ansatz results in only one single non-perturbative parameter 
$\alpha_0 = \mu_I^{-1}\int_0^{\mu_I} \alpha_\mathrm{eff}(k)\,\mathrm{d}k$,
being the first moment of the effective coupling integrated over the low scale region
up to the matching scale $\mu_I$.
Power corrections for event shape variables in DIS have been calculated to
one-loop~\cite{Dasgupta:1998ex} and two-loop~\cite{Dasgupta:1998xt} accuracy. 

For the differential distributions the power correction results in a shift of
the perturbatively calculated distribution~\cite{Dokshitzer:1997ew} 
\begin{equation}
  \frac{1}{\sigma}\frac{\diff\sigma(F)}{\diff F}=\frac{1}{\sigma}
                  \frac{{ \diff\sigma^{\mathrm{pQCD}}}(F-a_F\mathcal{ P})}{\diff F} \; ,
  \label{fshift}
\end{equation} 
where $a_F$ is of order one and can be calculated perturbatively. 
For the jet broadening a squeezing is applied in addition to
the shift, which is absorbed in the coefficient $a_F$~\cite{Dasgupta:2001eq}.

The power correction term $\mathcal{P}$ is assumed to be universal for all
event shape variables.
It is proportional to $1/Q$ and 
evaluated to be
\begin{equation}
  \mathcal{P}=\frac{16}{3\pi}\mathcal{M}\frac{\mu_I}{Q}
  \left [\alpha_0(\mu_I)-\alpha_s(Q)-\frac{\beta_0}{2\pi}
    \left(\ln \frac{Q}{\mu_I}+\frac{K}{\beta_0}+1\right)
    \alpha_s^2(Q) \right ] \; ,
  \label{pformula}
\end{equation}
where $\beta_0=11-2\,n_f/3$, 
$K = 67/6 -\pi^2/2-5\, n_f/9$, and
$n_f=5$ is the number of active flavours. 
The so-called Milan factor $\mathcal{M}\simeq1.05$
ensures the universality at the two-loop level~\cite{Dasgupta:1998xt}.
 
The simple shift in Eq.~\ref{fshift} cannot be valid over the whole spectrum.
At low values of $F$ it may be applied only for $F \gg a_F\mathcal{ P} \sim
\mu_I/Q$~\cite{Dokshitzer:1997ew}.
Moreover, at large values of $F$  higher order corrections are substantial and the NLO
calculation is not reliable.

The mean value of an event shape variable is modified through non-perturbative
effects by an additive constant~\cite{Dokshitzer:1995zt} 
\begin{equation}
  \mean{F} = { \mean{F}^{\textrm{\scriptsize pQCD}}}+a_F{ \mathcal{P}},
\end{equation}
with the same coefficient $a_F$ and the same function $\mathcal{P}$ as for the
distribution (Eq.~\ref{pformula}).

The theory predicts a universal value of $\alpha_0(\mu_I)$ of about 0.5~\cite{Dokshitzer:1995zt}.
For mean event shape values this prediction has been confirmed within 
$20\%$ in DIS~\cite{Adloff:1999gn} as well as in $\ee$~annihilation~\cite{Abdallah:2002xz}. 
Similar conclusions were drawn in an analysis of differential
distributions in $\ee$ annihilation~\cite{Abdallah:2002xz}.

\subsection{Fit Procedure}

Fits for $\alps$ and $\alpha_0$ are performed via a $\chi^2$ minimisation using MINUIT~\cite{James:1975dr},  
which for one event shape variable is defined as
\begin{equation}
  \chi^2=\sum_{i,j}\Delta_i V^{-1}_{ij} \Delta_j \, ,\quad \quad
  \Delta_i=m_i-t_i(\alps,\alpha_0) \, ,
  \label{eqchi2}
\end{equation}
with $V$ the covariance matrix, $i$ and $j$ extending over the bins 
included in the fit,
$m$ the measured data points and $t(\alps,\alpha_0)$ the theory
prediction, which depends on the free parameters. 
The covariance matrix $V$ 
consists of the sum of the individual covariance matrices from the different sources
presented in section~\ref{detcorrectionsection}:
\begin{equation}
  V=V_\textrm{stat}+V_\textrm{elm.escale}+V_\textrm{had.escale}
                   +V_\textrm{e.track}+V_\textrm{model}+V_\textrm{unfold}\, ,
\end{equation}
where $V_\textrm{stat}$ is determined by the unfolding procedure.
Only the unfolding and part of the model uncertainties are treated 
as uncorrelated, i.e.\ these matrices are diagonal.
Summing the covariance matrices in this way corresponds to adding the errors in quadrature.
The correlated part of the model uncertainty is determined as 
half of the difference in the fitted parameters when the fit is repeated,
using the program DJANGOH instead of RAPGAP for unfolding.

\subsection{Fits to Spectra}
\label{fitstospectra}
In order to fit the calculations described in section~\ref{qcdpheno} to the measured distributions,
the range of the distributions to be used has to be
specified.   
The upper bounds in $F$ are given in \cite{Dasgupta:2002dc} and are motivated by
properties of the perturbative calculations. 
The lower bounds in $F$ are set by the behaviour of the power correction, which limits
the reliability of the prediction at low $Q$, see Eqs.~\ref{fshift}-\ref{pformula}.
The present analysis makes an effort to extend the fit interval to values of
$F$ as low as possible.  

For a given event shape, the $\chi^2$  is first calculated using the
 highest $F$ and $Q$ bins only. Bins corresponding to lower $F$ values
 and lower scales are successively included provided that the  $\chi^2$
 of the fit does not increase by more than four units for each
 additional bin. 

The data are well described by the resummed
pQCD calculation supplemented by power corrections as shown
in Figs.~\ref{figureshapesdistributions1} and \ref{figureshapesdistributions2}.
In many cases even the lowest $F$ bin could be included in the fit,
which is reasonable since, although the distribution diverges as
$F \to 0$, the integral within the measured low $F$ bin remains finite.
At low $Q$ values the agreement between the measurements and the 
calculation degrades. In this domain the hadronisation effects 
become more important and the simple shift of Eq.~\ref{fshift} is not expected
to hold.

\begin{figure} [htb]
  \begin{center}
    \includegraphics[width=0.6\textwidth]{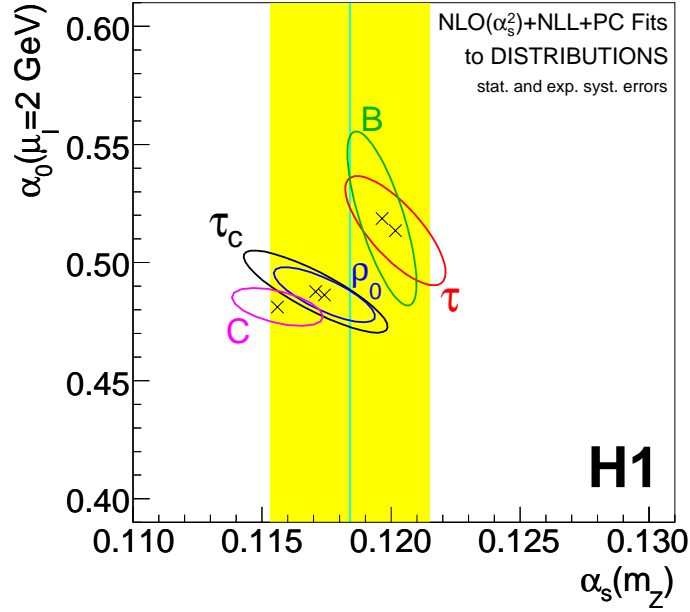}
  \end{center}
  \caption{Fit results to the differential distributions of
    $\tau$, $B$, $\rho_0$, $\tau_C$ and the $C$-parameter 
    in the $(\alpha_s,\alpha_0)$ plane.
    The $ 1\sigma$ contours
    correspond to $\chi^2 = \chi^2_{\rm min}+1$,
    including statistical and experimental systematic uncertainties.
    The value of $\alpha_s$ (vertical line) and its uncertainty (shaded band) are taken from \cite{Bethke:2004uy}. }
\label{figureellipsesdist}
\end{figure}

The results of the combined fit for $\alpha_0$ and $\alp(m_Z)$ are displayed in 
Fig.~\ref{figureellipsesdist} and summarised in Table~\ref{resulttable}.
The quality of the fits, expressed in terms of $\chi^2$ per degree of freedom,
is found to be reasonable.
For all event shape variables, consistent values for $\alp(m_Z)$ and $\alpha_0$ are found,
with a maximum difference of about two standard deviations between $\tau$ and $C$.
A strong negative correlation between  $\alp(m_Z)$ and  $\alpha_0$ is observed for all
variables. 
The values of the strong coupling $\alpha_s(m_Z)$ are in good agreement with
the world average~\cite{Bethke:2004uy}, shown for comparison as the
shaded band.
The non-perturbative parameter $\alpha_0\simeq 0.5$ is confirmed to be
universal within $10\%$.

\begin{table}[htb]
  \begin{center}
\renewcommand{\arraystretch}{1.15}
    \begin{tabular}{|l|c|c|c|c|c|}
\hline
\multicolumn{6}{|c|}{{\bf strong coupling constant} \boldmath $\alpsmz$\unboldmath }\\
\hline
 event shape variable&{$\tau_c$}&{$\tau$}&{$B$}&{$\rho_0$}&{$C$}\\
\hline
\hline
\hline
 central value&$ 0.1171 $&$ 0.1202 $&$ 0.1196 $&$ 0.1174 $&$ 0.1156 $\\
\hline
 uncertainties:&&&&&\\
\hline\multicolumn{1}{|r|}{total}&$^{+0.0068}_{-0.0062}$&$^{+0.0072}_{-0.0058}$&$^{+0.0072}_{-0.0064}$&$^{+0.0070}_{-0.0056}$&$^{+0.0073}_{-0.0054}$\\
\hline
\hline
\multicolumn{1}{|r|}{total experimental}&${\scriptstyle\pm0.0035}$&${\scriptstyle\pm0.0021}$&${\scriptstyle\pm0.0014}$&${\scriptstyle\pm0.0021}$&${\scriptstyle\pm0.0021}$\\
\hline
\multicolumn{1}{|r|}{ statistical experimental}&${\scriptstyle\pm0.0014}$&${\scriptstyle\pm0.0006}$&${\scriptstyle\pm0.0004}$&${\scriptstyle\pm0.0010}$&${\scriptstyle\pm0.0009}$\\
\hline
\multicolumn{1}{|r|}{ systematic experimental}&${\scriptstyle\pm0.0033}$&${\scriptstyle\pm0.0020}$&${\scriptstyle\pm0.0013}$&${\scriptstyle\pm0.0019}$&${\scriptstyle\pm0.0019}$\\
\hline
\hline
\multicolumn{1}{|r|}{total theoretical}&$^{+0.0058}_{-0.0051}$&$^{+0.0068}_{-0.0054}$&$^{+0.0071}_{-0.0063}$&$^{+0.0067}_{-0.0052}$&$^{+0.0069}_{-0.0049}$\\
\hline
\multicolumn{1}{|r|}{$\mu_r$ dependence }&$^{+0.0054}_{-0.0048}$&$^{+0.0058}_{-0.0043}$&$^{+0.0056}_{-0.0044}$&$^{+0.0064}_{-0.0050}$&$^{+0.0069}_{-0.0048}$\\
\hline
\multicolumn{1}{|r|}{$\mu_I$ dependence }&$^{+0.0002}_{-0.0002}$& $\scriptstyle<10^{-4}$& $\scriptstyle<10^{-4}$&$^{+0.0002}_{-0.0002}$& $\scriptstyle<10^{-4}$\\
\hline
\multicolumn{1}{|r|}{fit interval }&$^{+0.0015}_{-0.0018}$&$^{+0.0007}_{-0.0022}$&$^{+0.0001}_{-0.0009}$&$^{+0.0010}_{+0.0007}$&$^{+0.0003}_{-0.0004}$\\
\hline
\multicolumn{1}{|r|}{parton density functions }&$^{+0.0002}_{-0.0001}$&$^{+0.0003}_{-0.0010}$&$^{+0.0006}_{-0.0007}$&$^{+0.0001}_{-0.0002}$&$^{+0.0002}_{-0.0001}$\\
\hline
\multicolumn{1}{|r|}{matching scheme }&$^{+0.0015}_{+0.0005}$&$^{+0.0036}_{+0.0022}$&$^{+0.0043}_{+0.0043}$&$^{+0.0018}_{+0.0009}$&$^{-0.0005}_{-0.0009}$\\
\hline
\multicolumn{6}{c}{ \vspace{-0.2cm}}\\
\hline
\multicolumn{6}{|c|}{{\bf non perturbative coupling} \boldmath $\alpha_0(\mu_I=2\,\GeV)$\unboldmath }\\
\hline
 event shape variable&{$\tau_c$}&{$\tau$}&{$B$}&{$\rho_0$}&{$C$}\\
\hline
\hline
\hline
 central value&$ 0.488 $&$ 0.513 $&$ 0.519 $&$ 0.486 $&$ 0.481 $\\
\hline
 uncertainties:&&&&&\\
\hline\multicolumn{1}{|r|}{total}&$^{+0.037}_{-0.035}$&$^{+0.034}_{-0.039}$&$^{+0.059}_{-0.049}$&$^{+0.023}_{-0.035}$&$^{+0.028}_{-0.042}$\\
\hline
\hline
\multicolumn{1}{|r|}{total experimental}&${\scriptstyle\pm0.021}$&${\scriptstyle\pm0.025}$&${\scriptstyle\pm0.039}$&${\scriptstyle\pm0.014}$&${\scriptstyle\pm0.008}$\\
\hline
\multicolumn{1}{|r|}{ statistical experimental}&${\scriptstyle\pm0.009}$&${\scriptstyle\pm0.009}$&${\scriptstyle\pm0.006}$&${\scriptstyle\pm0.006}$&${\scriptstyle\pm0.005}$\\
\hline
\multicolumn{1}{|r|}{ systematic experimental}&${\scriptstyle\pm0.019}$&${\scriptstyle\pm0.023}$&${\scriptstyle\pm0.038}$&${\scriptstyle\pm0.013}$&${\scriptstyle\pm0.007}$\\
\hline
\hline
\multicolumn{1}{|r|}{total theoretical}&$^{+0.030}_{-0.027}$&$^{+0.022}_{-0.029}$&$^{+0.044}_{-0.029}$&$^{+0.019}_{-0.032}$&$^{+0.026}_{-0.041}$\\
\hline
\multicolumn{1}{|r|}{$\mu_r$ dependence }&$^{+0.020}_{-0.026}$&$^{+0.018}_{-0.027}$&$^{+0.030}_{-0.028}$&$^{+0.017}_{-0.027}$&$^{+0.022}_{-0.038}$\\
\hline
\multicolumn{1}{|r|}{fit interval }&$^{+0.022}_{-0.007}$&$^{+0.008}_{-0.005}$&$^{+0.030}_{+0.006}$&$^{-0.003}_{-0.016}$&$^{+0.006}_{-0.003}$\\
\hline
\multicolumn{1}{|r|}{parton density functions }&$^{+0.001}_{-0.001}$&$^{+0.006}_{+0.004}$&$^{+0.011}_{+0.003}$&$^{+0.001}_{-0.001}$&$^{+0.001}_{-0.002}$\\
\hline
\multicolumn{1}{|r|}{matching scheme }&$^{-0.005}_{-0.012}$&$^{-0.009}_{-0.023}$&$^{+0.006}_{-0.010}$&$^{-0.006}_{-0.009}$&$^{-0.014}_{-0.014}$\\
\hline
\multicolumn{6}{c}{ \vspace{-0.2cm}  }\\
\hline
correlation coefficient $\alp$,$\alpha_0$&$-0.85$&$-0.76$&$-0.75$&$-0.78$&$-0.51$\\
\hline
$\chi^2$ / d.o.f. (experimental errors)&$1.13$&$0.51$&$0.81$&$1.40$&$1.20$\\
\hline
\end{tabular}

    \caption{
        Results of simultaneous fits of $\alpha_s(m_Z)$ and $\alpha_0(\mu_I=2\,\GeV)$ to
        the distributions of the event shape variables $\tau_C$, $\tau$, $B$,
        $\rho_0$ and $C$.
        The statistical and experimental systematic errors as well as 
        the theoretical uncertainties are given. }
    \label{resulttable}
  \end{center}
\end{table}

The theoretical uncertainties on the fitted values of $\alpha_0$ and $\alp(m_Z)$
 are determined
from the changes to the results under variation in the procedure 
 as follows:
\begin{itemize}
  \item bins with lower boundaries at $F=0$ are omitted;
  \item the renormalisation scale $\mu_r$ is varied from $Q/2$ to $2Q$;
  \item the infrared matching scale $\mu_I$ is varied from  $1.5\,\GeV$ to $2.5\,\GeV$;
  \item the CTEQ proton pdfs are replaced by three versions of the MRST2001 set~\cite{Martin:2001es}, 
    which differ in $\alps(m_Z)$ from $0.117$ to $0.121$;
  \item instead of $\log R$ the  modified $M$ and modified $M^2$ matching schemes \cite{Dasgupta:2002dc} are used.
\end{itemize}
The fit procedure is repeated for each of these variations separately.

All fit results, including the individual contributions to the total error, are
given in numerical form in Table~\ref{resulttable}. 
The theoretical error is the dominant contribution to
the total uncertainty and arises mainly due to the renormalisation scale uncertainty. 

The good agreement of the results for all event shape variables allows
a common set of values of $\alps(m_Z)$ and $\alpha_0$ to be derived
by applying an averaging procedure to the results from the individual
event shape variables. In this procedure, the
$\chi^2$ minimisation takes into account all experimental and theoretical
errors and the correlations between $\alps(m_Z)$ and $\alpha_0$ as given in Table~\ref{resulttable}.
In addition the correlations among the observables are considered.
The precision on the very large correlation coefficients
 within the group \{$\tau,\,B$\} 
and the group \{$\tau_C,\,\rho_o,\,C$\}
is not sufficient to allow the correlation
matrix to be used directly in a $\chi^2$ minimisation.
Instead the averaging is performed in two steps. 
Firstly, the results within these two groups are combined, neglecting the correlations
among the observables.
The smallest uncertainty of the contributing measurements is conservatively taken as the
uncertainty of the group average. 
Secondly, the two group averages, being only moderately correlated,
are combined using the corresponding correlation matrix.

The averaging procedure results in:
\begin{eqnarray}
\label{combaseqn}
  \alps(m_Z) & = &
      0.1198 \pm 0.0013\ ({\rm exp})\ ^{+0.0056} _{-0.0043}\ ({\rm theo}) \ ,
       \nonumber 
       \\[-1ex]  &  &  \label{eq_res2parm}  \\[-1ex]
  \alpha_0   & = &
      0.476 \pm  0.008 \ ({\rm exp})\ ^{+0.018} _{-0.059}\  ({\rm theo}) \ ,
      \nonumber
\end{eqnarray}
with a fit quality of $\chi^2/{\rm d.o.f.} = 4.9/2$.
Here  the theoretical error is derived from the renormalisation scale uncertainty.
Note that the combined value of $\alpha_0$ is lower than 
the individual values, due to the negative correlations between
 $\alpha_0$ and $\alps(m_Z)$. 

If instead the correlations between the two groups of event shape observables are neglected, 
 a consistent result is obtained.

\subsection[Running of Strong Coupling $\alpha_s(Q)$ ]{Running of Strong Coupling \boldmath $\alpha_s(Q)$ }

In the previous section it was shown that the concept of power
corrections provides a good description of hadronisation effects in the
differential event shape distributions. 
Alternatively, one may
assume the validity of the power corrections and
 investigate the scale dependence of the strong coupling 
$\alpha_s(Q)$.
For each event shape variable a $Q$ independent $\alpha_0$ parameter
and an $\alps(Q)$ for each $Q$ bin are fitted.

\begin{figure} [htb]
  \begin{center}
    \includegraphics[width=0.75\textwidth]{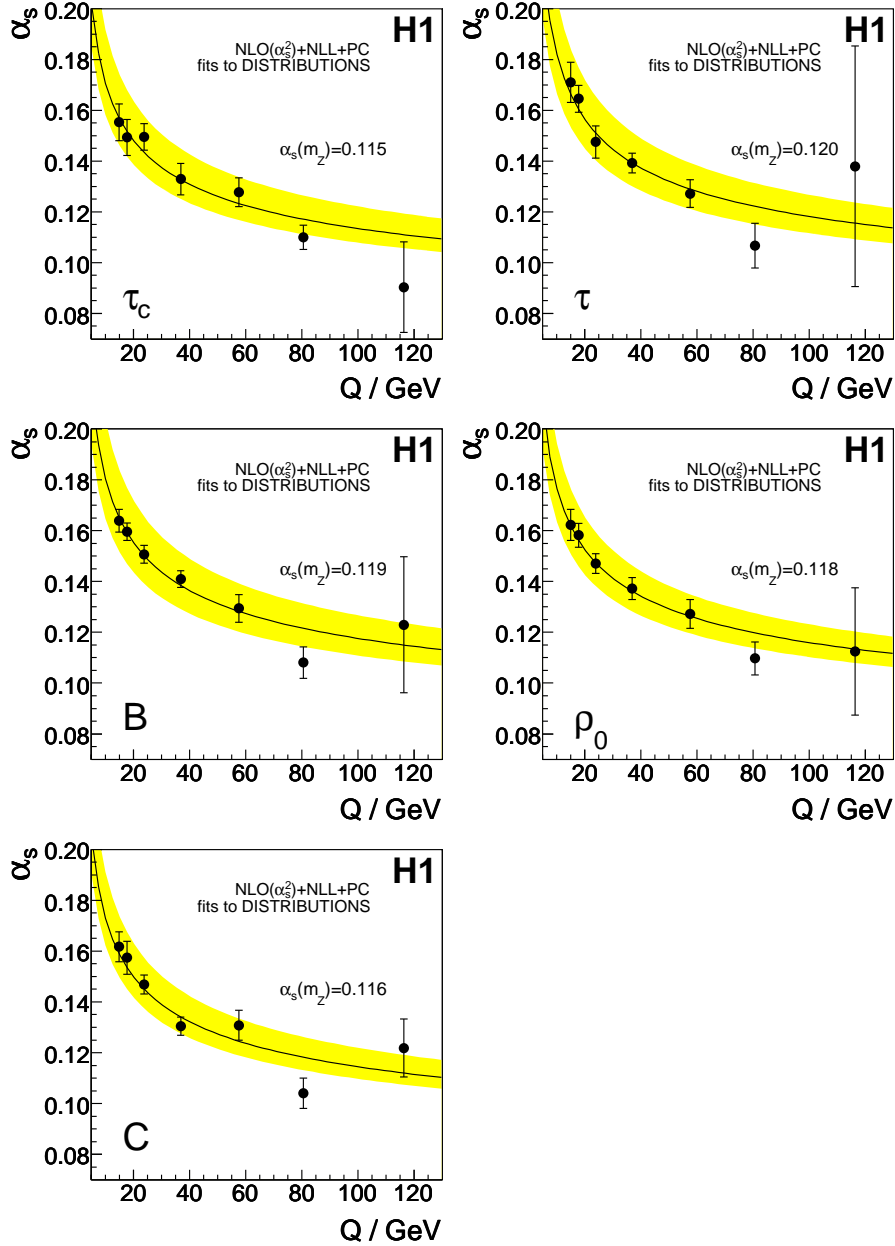}
  \end{center} 
  \caption{
The strong coupling $\alps$ as a function of the scale $Q$.
The individual fit results, shown
as points with error bars, are obtained
from fits to the differential distributions in $\tau_C$, $\tau$, $B$, $\rho_0$ and $C$
within each $Q$ bin.
The errors represent the total experimental uncertainties.
For each event shape observable a value of 
$\alps(m_Z)$ is indicated in the plot, determined
from a fit to the $\alps(Q)$ results using the QCD renormalisation
group equation.
The corresponding fit curves are shown as full lines.
The shaded bands represent the uncertainties on $\alps(Q)$ 
from renormalisation scale variations.}
  \label{runningas} 
\end{figure}

The fitted values of $\alps(Q)$ are presented in 
Fig.~\ref{runningas}.
The running of the strong coupling 
 is clearly observed for each of the event shape variables
over a wide scale range between $\langle Q \rangle =15\,\GeV$  and $\langle Q \rangle =116\,\GeV$.
The numerical values with experimental errors are given in Table~\ref{runningastable}.
The theoretical uncertainties are of similar size to those given in Table~\ref{resulttable}.

\begin{table}[htb]
  \begin{center}
    \small \footnotesize\begin{tabular}{|l||c|c|c|c|c|c|}
\hline
 $Q/$&\multicolumn{6}{|c|}{ {\bf strong coupling constant} \boldmath $\alps(Q)$\unboldmath }\\
\cline{2-7}
GeV&{$\tau_c$}&{$\tau$}&{$B$}&{$\rho_0$}&{$C$}&{$\langle \alps(Q) \rangle$}\\
\hline
\hline
$ 15$ &$0.155\pm 0.007$ &$0.171\pm 0.008$ &$0.164\pm 0.004$ &$0.162\pm 0.006$ &$0.162\pm 0.006$ &$0.163\pm0.004$ \\
$ 18$ &$0.149\pm 0.007$ &$0.165\pm 0.005$ &$0.160\pm 0.003$ &$0.158\pm 0.005$ &$0.157\pm 0.006$ &$0.159\pm0.003$ \\
$ 24$ &$0.150\pm 0.005$ &$0.148\pm 0.006$ &$0.151\pm 0.003$ &$0.147\pm 0.004$ &$0.147\pm 0.004$ &$0.148\pm0.003$ \\
$ 37$ &$0.133\pm 0.006$ &$0.139\pm 0.004$ &$0.141\pm 0.003$ &$0.137\pm 0.004$ &$0.130\pm 0.004$ &$0.136\pm0.003$ \\
$ 58$ &$0.128\pm 0.006$ &$0.127\pm 0.005$ &$0.129\pm 0.005$ &$0.127\pm 0.006$ &$0.131\pm 0.006$ &$0.128\pm0.005$ \\
$ 81$ &$0.110\pm 0.005$ &$0.107\pm 0.009$ &$0.108\pm 0.006$ &$0.110\pm 0.006$ &$0.104\pm 0.006$ &$0.108\pm0.004$ \\
$116$ &$0.090\pm 0.018$ &$0.138\pm 0.047$ &$0.123\pm 0.027$ &$0.112\pm 0.025$ &$0.122\pm 0.011$ &$0.114\pm0.011$ \\
\hline
\hline
 $\alpha_0$ &$0.496\pm 0.014$ &$0.515\pm 0.020$ &$0.516\pm 0.034$ &$0.484\pm 0.011$ &$0.477\pm 0.007$ & \\
\hline
\end{tabular}

    \caption{Fitted values of $\alps(Q)$ as determined from the 
    differential distributions of the event shape variables
    $\tau_C$, $\tau$, $B$, $\rho_0$ and $C$ and the averaged values $\langle \alps(Q) \rangle$ ;
    the fitted parameters $\alpha_0$ are given in the last line.}
    \label{runningastable}
  \end{center}
\end{table}

\begin{figure} [htb]
  \begin{center}
    \includegraphics[width=0.6\textwidth]{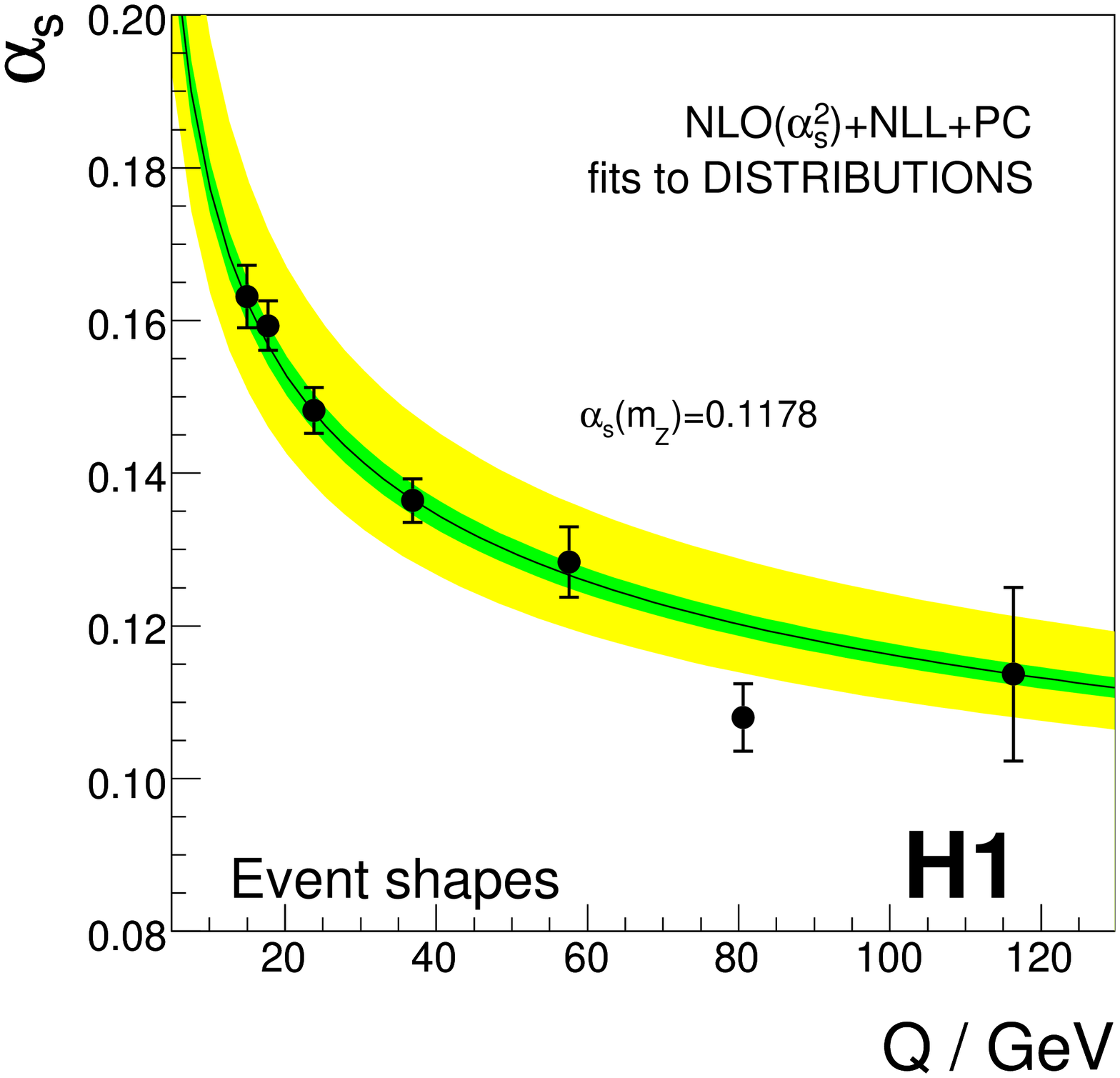}
  \end{center} 
  \caption{ The strong coupling $\alps$ as a function of the scale $Q$ from 
    an average of the results obtained by fitting the
differential event shape distributions.
The errors represent the total experimental uncertainties.
A value of $\alps(m_Z)$ is indicated in the plot, determined from a fit to the  $\alps(Q)$ results using the
QCD renormalisation group equation.
The fit curve is shown as the full line.
The inner (outer) shaded band represents the uncertainty of the fitted 
$\alps(Q)$ from experimental errors (the renormalisation scale variation).
  }
  \label{runningascomb} 
\end{figure} 

The fit results of the different event shapes are compatible with each
other and again may be combined.
The correlations among the observables are taken into account in a two-step
procedure as described in the previous section. 
The averaged $\alps(Q)$ values  are displayed as a function of $Q$ in
Fig.~\ref{runningascomb} and listed in Table~\ref{runningastable}.
A fit of the renormalisation group equation to the measured
 $\alps(Q)$  yields 
\begin{eqnarray}
  \alps(m_Z) & = &
      0.1178 \pm 0.0015 \ ({\rm exp})\ ^{+0.0081} _{-0.0061}\ ({\rm theo}),
  \label{eq_res1parm} 
\end{eqnarray}
with $\chi^2/{\rm ndf} = 8.3/6$.
This value is in good agreement with the two-parameter fit result quoted in
Eq.~\ref{eq_res2parm}, though the scale uncertainty is somewhat larger here.
The difference with respect to the result quoted in Eq.~\ref{combaseqn} is that
here individual power correction parameters $\alpha_0$
are associated to each observable.

\subsection{Fits to Mean Values}

The mean values of the event shape variables,
presented as a function of the scale $Q$ in Fig.~\ref{distrmeans}, 
are also subjected to QCD fits. 
For this application the resummed calculation can not be used,
because of difficulties at high values of the event shape variable, which are
related to sub-leading logarithms and the matching procedure \cite{privsalam}. 
These regions are excluded from the fits to the full spectra, but by definition contribute
to the mean values. 
Hence the theoretical prediction for the mean values is solely based on a NLO
calculation, supplemented by power corrections.

\begin{figure} 
  \begin{center}
    \includegraphics[width=0.5\textwidth]{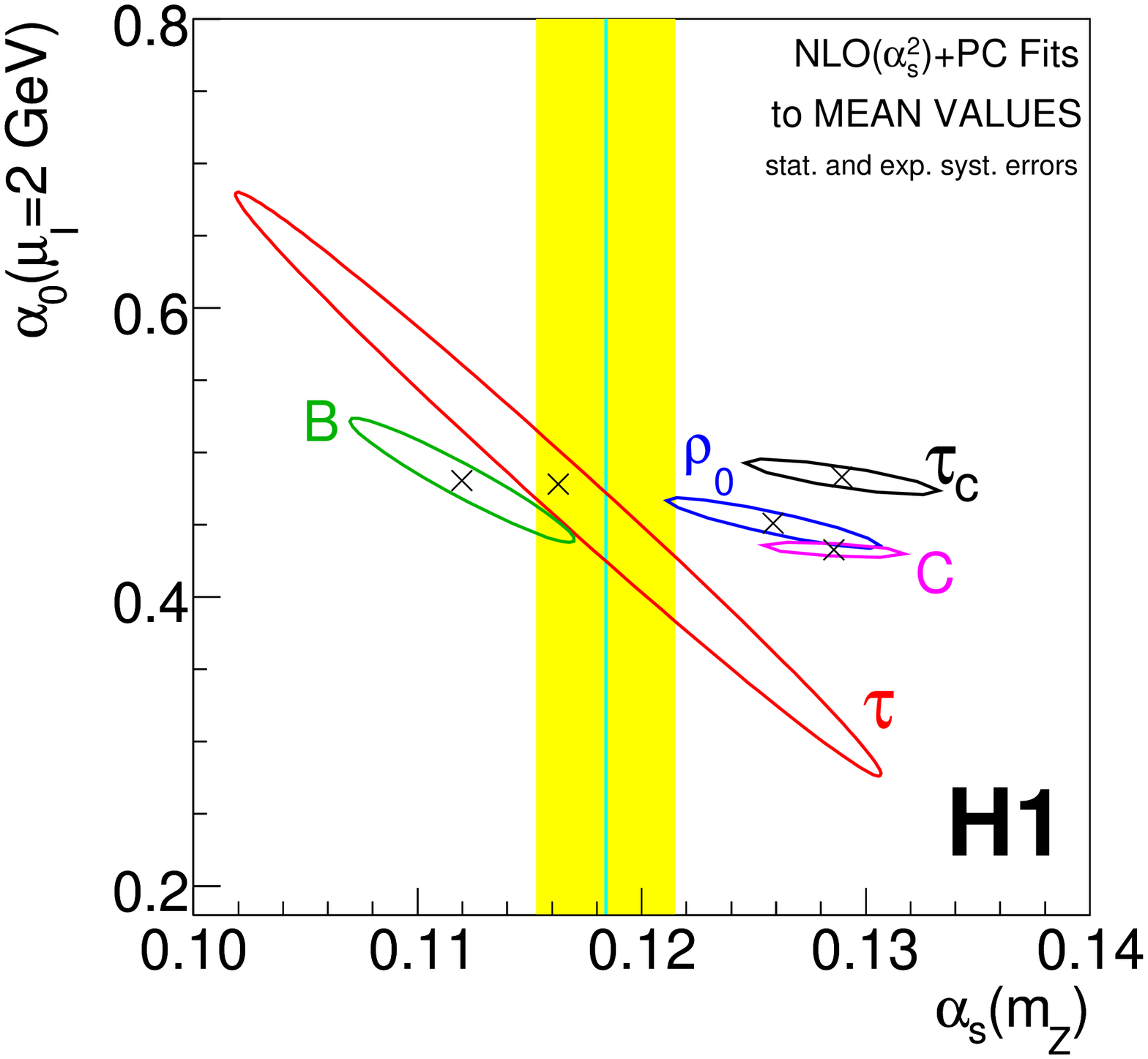}
  \end{center} 
  \caption{Results of fits to the mean values of
    $\tau$, $B$, $\rho_0$, $\tau_C$ and the $C$-parameter 
    in the $(\alpha_s,\alpha_0)$ plane.
    The $ 1\sigma$ contours
    correspond to $\chi^2 = \chi^2_{\rm min}+1$,
    including statistical and experimental systematic uncertainties.
    The value of $\alpha_s$ (vertical line) and its uncertainty (shaded band) are taken from \cite{Bethke:2004uy}.
    Note the enlarged scale compared to Fig.~\ref{figureellipsesdist}.
}
\label{figureellipsesmean} 
\end{figure}

The results of the fits are displayed in Fig.~\ref{distrmeans} and the fitted
values of $\alpha_s(m_Z)$ and $\alpha_0$ are shown in Fig.~\ref{figureellipsesmean}.
The non-perturbative parameters cluster around a common value
of $\alpha_0\simeq 0.45 - 0.50$, thus supporting universality 
at the 10\% level. 
The fitted values for $\alpha_s(m_Z)$ exhibit a rather large spread.

Compared to the results obtained from the event shape distributions shown in Fig.~\ref{figureellipsesdist},
higher values of $\alps$ are found using $\tau_C$, $\rho_0$ and in particular the $C$-parameter,
and a somewhat lower value is obtained from $B$.
The analyses of both the event shape distributions and mean values
are based on the same data and there is no experimental reason
why they should lead to different results.
Therefore, it is likely that the different theoretical treatment for mean values
is the source of the observed deviations.

A dedicated study of mean values has been published previously by the H1
collaboration~\cite{Adloff:1999gn} for $7 < Q < 100\,\GeV$. 
As mentioned in section~\ref{resultssec} the current and previous measurements
 are in agreement in the phase space of overlap.
The results of both QCD analyses are also consistent with each other. 
The previous analysis shows a larger sensitivity to the parameters  
$\alps$ and $\alpha_0$ because it includes data at lower $Q$.
This leads to substantially reduced error ellipses compared to those of 
Fig.~\ref{figureellipsesmean},
in particular for the $\langle \tau \rangle$ observable.

Differences between sets of QCD parameters $\alps$ and $\alpha_0$
determined in differential distributions and mean values,
similar to those reported here, 
have been observed in $e^+e^-$ annihilation by DELPHI~\cite{Abdallah:2002xz}. 
Discrepancies have also been found by other experiments 
\cite{MovillaFernandez:2001ed,Heister:2003aj}.
In general, similar values of $\alpha_0\simeq 0.5$ 
are found for event shape variables in \ee scattering, but with a much larger 
spread than in deep-inelastic scattering,
the jet broadening results in particular being different from the others.

\section{Conclusions}

Accurate measurements of event shape variables in
deep-inelastic $ep$ scattering are presented based on  $106~\pb^{-1}$ of data
with four-momentum transfer $Q$ ranging between  $14~\GeV$ and $200~\GeV$.
Resummed perturbative QCD predictions together with power corrections
give good descriptions of the spectra of the observables thrust,
jet broadening, jet mass and $C$-parameter.
The use of resummed calculations extends the good
description to low values of the event shape variables, corresponding to pencil like 
configurations. 

The results of a two-parameter fit of the strong coupling constant
$\alp$ and the effective non-perturbative coupling $\alpha_0$
for the various event shape observables are consistent with each other. 
The values for $\alp$ agree with the world average.
The parameter $\alpha_0$, which accounts for hadronisation, is 
consistently found to be $0.5$ within $10\%$,
in good agreement with theoretical expectation.
A combined analysis of all event shape variables yields
\begin{eqnarray}
  \alps(m_Z) & = &
      0.1198 \pm 0.0013\ ({\rm exp})\ ^{+0.0056} _{-0.0043}\ ({\rm theo}) \ ,
      \nonumber \\[1ex]
  \alpha_0   & = &
      0.476 \pm  0.008 \ ({\rm exp})\ ^{+0.018} _{-0.059}\  ({\rm theo}) \ ,
      \nonumber
\end{eqnarray}
where the theoretical error is derived from the renormalisation scale uncertainty.
Relaxing the requirement of a common value of $ \alpha_0$,
the data are used to
investigate the scale dependence of the strong coupling over a wide range
of $Q = 15 - 116~\GeV$. 
The running of $\alps(Q)$ is clearly observed for 
each event shape variable, in accordance with the expected evolution.
Combining the results of all variables leads to
$\alps(m_Z) =
      0.1178 \pm 0.0015 \ ({\rm exp})\ ^{+0.0081} _{-0.0061}\ ({\rm theo})$.
The errors are dominated by the renormalisation scale uncertainty, which suggests
that missing higher order terms in the perturbative calculation are important.

In the analysis of the event shape means the results for $\alps(m_Z)$ and $\alpha_0$
are less accurate than those obtained in the fits to the distributions.
While the non-perturbative parameters $\alpha_0$ cluster again around a common value
of $0.5$, the results for the strong coupling $\alp$ obtained from the 
five event shape means exhibit a spread considerably larger
than is expected from their individual uncertainties. 
These discrepancies
may be related to an insufficient theoretical treatment, which lacks resummed calculations.

The observed universality of $\alpha_0$ for both
 distributions and mean values of the event shape variables supports the
concept of power corrections.
Hence, it can be considered as an appropriate alternative
to conventional models for the description
of hadronisation effects for event shape variables.

\section*{Acknowledgments}

We are grateful to the HERA machine group whose outstanding
efforts have made this experiment possible. 
We thank
the engineers and technicians for their work in constructing and
maintaining the H1 detector, our funding agencies for 
financial support, the
DESY technical staff for continual assistance
and the DESY directorate for support and for the
hospitality which they extend to the non DESY 
members of the collaboration.
We want to thank M.~Dasgupta and G.P.~Salam for many valuable discussions and
for good cooperation over the years.


\end{document}